\documentclass[12pt,graphicx]{article}
\usepackage{amssymb,amsmath,amsfonts}
\usepackage{graphicx}
\usepackage{graphics}
\usepackage{eepic,epsfig}
\newcommand{\n}{{\not\hspace{-0.5ex}\nabla}}
\newcommand{\D}{{\not\hspace{-0.8ex}D}}
\begin{document}
\title{Fermionic vacuum polarization in higher-dimensional global monopole spacetime}
\author{E. R. Bezerra de Mello \thanks{E-mail: emello@fisica.ufpb.br}\\
Departamento de F\'{\i}sica-CCEN\\
Universidade Federal da Para\'{\i}ba\\
58.059-970, J. Pessoa, PB\\
C. Postal 5.008\\
Brazil}
\maketitle

\begin{abstract} In this paper we analyse the vacuum polarization effects associated with a massless fermionic field in a higher-dimensional global monopole spacetime in the "braneworld" scenario. In this context we admit that the our Universe, the bulk, is represented by a flat $(n-1)-$dimensional brane having a global monopole in a extra transverse three dimensional submanifold. We explicitly calculate the renormalized vacuum average of the energy-momentum tensor, $\langle T_A^B(x)\rangle_{Ren.}$, admitting the global monopole as being a point-like object. We observe that this quantity depends crucially on the value of $n$, and we provide explicit expressions to it for specific values attributed to $n$.
\end{abstract}

\newpage
\renewcommand{\thesection}{\arabic{section}.}
\section{Introduction}
Recently the braneworld model has attracted renewed interest. By this scenario our world is represented by a four dimensional sub-manifold, a three-brane, embedded in a higher dimensional spacetime \cite{Akama}. Braneworlds naturally appear in the string/M theory context and provide a novel setting for discussing phenomenological and cosmological issues related to extra dimensions. The models introduced by Randall and Sundrum are particularly attractive \cite{RS,RS1}. The corresponding spacetime contains two (RSI), respectively one (RSII), Ricci-flat brane(s) embedded on a five-dimensional Anti-de Sitter (AdS) bulk. It is assumed that all matter fields are confined on the branes and only the gravity propagates in the five dimensional bulk. The idea that matter is confined to a lower dimensional manifold is not a new one. The localization of fermions on a  domain wall has been discussed in \cite{R-S}. 

The hierarchy problem between the Planck scale and the electroweak one is solved in the RS1 model, if the distance between the two branes is about $37$ times the AdS radius. The braneworld model also provide some alternative discussions about one of the most important problem in the modern physics: the cosmological constant problem (see, for instance, Ref. \cite{CP}). In this way, the Casimir energy associated with quantum fields which propagates in the bulk obeying specific boundary conditions on the branes may contribute to both, the brane and bulk cosmological constant. The Casimir energy associated with scalar field on the five-dimensional Randal-Sundrum model are calculated in \cite{NO}. Surface Casimir densities and induced cosmological constant on the branes are calculated in \cite{Sah} for a massive scalar quantum field obeying Robin boundary conditions on two parallel brane in a general $(D+1)-$dimensional anti-de Sitter bulk.

Although topological defects have been first studied in a four-dimensional spacetime \cite{VS}, they have been considered in spacetimes of higher dimensions in the context of braneworld. In this scenario the defects live in a $d-$dimensions submanifold, having their cores on the $3-$brane. In this way, domain wall and cosmic string cases have been analyzed in \cite{R-S} and \cite{Cohen,Ruth}, respectively, considering $d=1$ and $2$. Local and global monopoles have also been analyzed in \cite{Roessl,Cho} and \cite{Ola} to \cite{Cho2}, respectively, considering $d=3$. Specifically in \cite{Cho2} it was shown that if $\eta_0$, the energy scale where the gauge symmetry of the global system is spontaneously broken, is smaller than the Planck mass, the seven-dimensional Einstein equations admit a solution which for points outside the global monopole's core can be expressed by
\begin{eqnarray}
\label{g}
	ds^2=\eta_{\mu\nu}dx^\mu dx^\nu+\frac{dr^2}{\alpha^2}+r^2(d\theta^2+\sin^2\theta d\phi^2)=g_{AB}d^A dx^B \ ,
\end{eqnarray}
where $\eta_{\mu\nu}=diag(-1 \  , \ 1 \ , \ ... \ , \ 1)$ is the Minkowski metric, $\alpha^2= 1-\kappa^2\eta^2_0$, a parameter smaller than unity, and $\kappa$ related with the seven-dimensional Planck mass. In order to be more precise, in \cite{Ola} the authors have obtained the solution to the Einstein equations considering a general $n-$dimensional Minkowski brane worldsheet and a $d\geq 3$ global monopole in the transverse extra dimensions. In this general case the metric is a generalization of (\ref{g}) with $\mu$ , $\nu= \ 0 \ , 1 \ ... \ , \ n-1$ and $\alpha^2=1-\frac{\kappa^2\eta^2_0}{d-2}$. The solid angle associated with the $3-$geometry above depends on the parameter $\alpha$ and reads $\Omega=4\pi^2\alpha^2$, so smaller than the usual one. Consequently in this submanifold there is a solid angle deficit $\Delta\Omega=4\pi^2\kappa^2\eta_0^2$ 

Composite topological defects have also been first studied in a four-dimensional space-time. Specifically a composite monopole, i.e., a system composed by a local and a global monopoles was analyzed in  \cite{Mello,BH,Mello1} (for composite strings see \cite{Mello2}). More recently, the composite monopole has been analysed in the braneworld scenario in \cite{Mello3}.

In previous publication \cite{Mello4}, we have analyzed the vacuum polarization effects associated with a massless scalar quantum field in an $(n+3)-$dimensional bulk spacetime which has the structure of a $n-$dimensional Minkowiski brane with a global monopole in the transverse three-dimensional sub-manifold.\footnote{Although the physical interesting case corresponds to $n=4$, we developed our formalism considering $n$ as an arbitrary number} Specifically we calculated the renormalized vacuum expectation value of the square of the field, $\langle\Phi^2(x)\rangle_{Ren}$, and have showed that this quantity depends crucially on the values attribute to $n$. We also analyzed the structure of the renormalized vacuum expectation value of the energy-momentum tensor, $\langle T_{AB}(x)\rangle_{Ren.}$. In order to develop these investigations we calculated the respective Euclidean scalar Green function, $G_E^{(n)}(x,x')$. 

Continuing in the same line of investigation, in this paper we shall analyze the polarization effect of fermionic vacuum induced by a point-like three-dimensional global monopole embedded in a higher-dimensional bulk in a braneworld scenario. Specifically we shall consider that the bulk has its geometric structure given by the line element (\ref{g}). Our main objective is to calculate the renormalized vacuum expectation value of the energy-momentum tensor, $\langle T_{AB}(x)\rangle_{Ren.}$. Differently from the scalar case, here we obtain a simpler expression to the fermionic Green function and for the energy-momentum tensor, which is obtained in a closed form for an arbitrary parameter angle deficit.

This paper is organized as follows: In section $2$ we calculate the Euclidean fermionic propagator associated with a massless field in the background of a point-like global monopole transverse to a flat $n-$dimensional brane in a braneworld scenario, considering $n=1, \ 2$ and $3$. In order to do that we write the general Dirac differential operator and the equation obeyed by the propagator. In section $3$ we analyze this function in the coincidence limit and extract all divergencies from it in manifest form. In this way we provide the explicit expressions to the components of the renormalized energy-momentum tensor, and analyze their behavior in some limiting cases. In section $4$, we summarize our most important results. In this paper we use signature $+2$, and the definitions: $R_{\beta\gamma\delta}^\alpha=\partial_\gamma\Gamma_{\beta\delta}^\alpha-...$, $R_{\alpha\beta}= R_{\alpha\gamma\beta}^\gamma$. We also use untis $\hbar=c=1$.

\section{Spinor Green Function}
\label{Spinor}
Before to start the calculation of the spinor Green function in the six-dimensional global monopole spacetime, we shall first review briefly some important properties of the Dirac equation in a flat space.

In a flat six-dimensional space, the Dirac matrices, $\Gamma^{(M)}$, are $8\times 8$ matrices, which can be construct from the four-dimensional $4\times 4$ ones \cite{B-D} as shown below \cite{Moha}:
\begin{eqnarray}
\Gamma^{(\mu)}=\left( 
\begin{array}{cc}
0&\gamma^\mu \\
\gamma^\mu&0 
\end{array} \right) \ , \
\Gamma^{(4)}=\left( 
\begin{array}{cc}
0&i\gamma_5\\
i\gamma_5&0 
\end{array} \right) \ , \
\Gamma^{(5)}=\left( 
\begin{array}{cc}
0&I \\
-I&0 
\end{array} \right) \ ,
\end{eqnarray}
where  $\gamma_5=i\gamma^o\gamma^1\gamma^2\gamma^3$, and $I$ the $4\times 4$ identity matrix. It can be easily verified that these matrices obey the Clifford algebra: $\{\Gamma^{(M)}, \ \Gamma^{(N)}\}=-2\eta^{(M)(N)}$, for $M, \ N=0, \ 1, \ ... \ 5$. 

In these representation, the $\Gamma^{(7)}$ matrix is written as:
\begin{eqnarray}
	\Gamma^{(7)}=\Gamma^{(0)}\Gamma^{(1)} \ ... \Gamma^{(5)}=\left( 
\begin{array}{cc}
I&0 \\
0&-I
\end{array} \right) \ .
\end{eqnarray}
This matrix has two chiral eigenstate defined as $\Psi_+$ and $\Psi_-$. Consequently any six-dimensional fermionic wave-function, $\Psi$, can be decomposed in terms of its chiral components as\footnote{It is worth to mention that the six-dimensional chirality does not correspond to four-dimensional chirality. Six-dimensional chiral wave-functions correspond to a four-dimensional wave-functions, which still contain two four-dimensional chiral components eigenstates of $\gamma^5$.} 
\begin{eqnarray}
\Psi=\left( 
\begin{array}{cc}
\Psi_+\\
\Psi_- 
\end{array} \right) \ .
\end{eqnarray}

Solutions for the Dirac equation,
\begin{eqnarray}
\label{D}
	i\Gamma^{(M)}\partial_{(M)}\Psi=M\Psi \ , 
\end{eqnarray}
with defined chirality can only be possible for $M=0$. In this way for positive chirality, equation (\ref{D}) reduces to
\begin{eqnarray}
	\sigma^{(M)}\partial_{(M)}\Psi_+=0 \ , 
\end{eqnarray}
being $\sigma^{(M)}=(\gamma^\mu, \ i\gamma_5, \ -I)$ a set of $4\times 4$ matrix; and for negative chirality it reduces to
\begin{eqnarray}
	{{\tilde{\sigma}}}^{(M)}\partial_{(M)}\Psi_-=0 \ , 
\end{eqnarray}
being now ${\tilde{\sigma}}^{(M)}=(\gamma^\mu, \ i\gamma_5, \ I)$.

In order to write the Dirac equation in the six-dimensional global monopole spacetime, we shall choose the following coordinate system and basis tetrad:
\begin{eqnarray}
\label{coord}
	x^A=(t, \ r , \theta, \ \phi, \ x, \ y)=(x^\mu, \ y^a) \ ,
\end{eqnarray}
where $\mu=0, \ 1, \ 2, \ 3$ and $a=1, \ 2$, and 
\begin{eqnarray}
\label{e}
	e^A_{(M)} = \left( \begin{array}{cccccc}
  1&0&0&0&0&0 \\
  0&\alpha\sin\theta \cos\phi & \cos \theta\cos\phi /r & -\sin\phi /r\sin\theta&0&0 \\ 
0&\alpha\sin\theta \sin\phi & \cos \theta\sin\phi /r & \cos\phi /r\sin\theta &0&0\\ 
0 & \alpha\cos\theta & -\sin \theta /r & 0&0&0\\
0&0&0& 0&1&0\\
0&0&0&0&0&1
                      \end{array}
               \right) \ .
\label{Tetrada}
\end{eqnarray}

The Dirac equation for a massive field in the above coordinate system reads
\begin{eqnarray}
\label{D1}
i\n\Psi-M\Psi=0 \ ,	
\end{eqnarray}
with the covariant derivative operator given by
\begin{eqnarray}
	\n=e^A_{(M)}\Gamma^{(M)}(\partial_A+\Pi_A) \ ,
\end{eqnarray}
where $\Pi_A$ is the spin connection, given in terms of the flat spacetime Dirac matrices by
\begin{eqnarray}
	\Pi_A=-\frac14\Gamma^{(M)}\Gamma^{(N)}e^C_{(M)}e_{(N)C;A} \ .
\end{eqnarray}

For the above basis tetrad, the only nonzero spin connections are:
\begin{eqnarray}
	\Pi_\theta&=&\frac{i}2(1-\alpha){\vec{\Sigma}}_{(8)}\cdot{\hat{\phi}} \nonumber\\
	\Pi_\phi&=&-\frac{i}2(1-\alpha)\sin\theta{\vec{\Sigma}}_{(8)}\cdot{\hat{\theta}} \ ,
\end{eqnarray}
where
\begin{eqnarray}
{\vec{\Sigma}}_{(8)}=\left( 
\begin{array}{cc}
{\vec{\Sigma}}&0 \\
0&{\vec{\Sigma}}
\end{array} \right) \ , \
{\vec{\Sigma}}=\left( 
\begin{array}{cc}
{\vec{\sigma}}&0\\
0&{\vec{\sigma}} 
\end{array} \right)  \ ,
\end{eqnarray}      
being ${\hat{\theta}}$ and ${\hat{\phi}}$ the standard unit vectors along the angular directions and $\sigma^k$ the Pauli matrices.             
  
The fermionic propagator obeys the following differential equation:
\begin{eqnarray}
	\left(i\n-M\right)S_F(x,x')=\frac1{\sqrt{-g}}\delta^{(6)}(x-x')I_{(8)} \ , 
\end{eqnarray}
where $g=det(g_{AB})$ and $I_{(8)}$ is the $8\times 8$ identity matrix. This propagator is a bispinor, i.e., it transforms as $\Psi$ at point $x$ and as ${\bar{\Psi}}$ at $x'$.

If a bispinor $D_F(x',x)$ satisfies the differential equation below
\begin{eqnarray}
	\left(\Box-M^2-{1\over 4}R\right)D_F(x,x') = -{1\over\sqrt{-g}} \delta^{(6)} (x-x')I_{(8)} \ ,
\label{GF0}
\end{eqnarray}
with generalized d'Alembertian operator given by
\begin{eqnarray}
	\Box=g^{MN}\nabla_M\nabla_M= g^{MN}\left(\partial_M \nabla_N + \Pi_M \nabla_N-\{^S_{MN}\} \nabla_S\right) \ ,  
\end{eqnarray}
the spinor Feynman propagator can be written as
\begin{eqnarray}
	S_F(x',x)=(i\n+M)D_F(x',x) \ .
\end{eqnarray}

Now, after this brief review about the calculation of spinor Feynman propagator, let considering the six-dimensional global monopole spacetime (\ref{g}), where the scalar curvature $R=\frac{2(1-\alpha^2)}{r^2}$. Choosing the basis tetrad (\ref{e}), we obtain, after some intermediate steps, that 
\begin{eqnarray}
\label{K0}	{\cal{K}}=\Box-\frac14R=-\partial_t^2+\alpha^2\left(\partial_r^2+\frac2r\partial_r\right)-\frac{{\vec{L}}^2}{r^2}+\partial_x^2+\partial_y^2-\frac{(1- \alpha)}{r^2}\left(1+{\vec{\Sigma}}_{(8)}\cdot{\vec{L}}\right) \ ,
\end{eqnarray}
being $\vec{L}$ the ordinary angular momentum operator.
                    
The system that we shall consider consists of a massless positive chiral fields. In this way Eq. (\ref{D1}) can be written in terms of a $4\times 4$ matrix differential equation
\begin{eqnarray}
	\D\Psi_+=0 \ ,  
\end{eqnarray}
with
\begin{eqnarray}
\label{DO}
	\D=\gamma^0\partial_t+\alpha\gamma_r\partial_r-\frac1r\gamma_r\left({\vec{\Sigma}}\cdot{\vec{L}}+1\right)+i\gamma_5\partial_x-I\partial_y +\frac\alpha r\gamma_r\ ,
\end{eqnarray}
being $\gamma_r=\hat{r}\cdot\vec{\gamma}$.

The Feynman four-component propagator obeys the equation
\begin{eqnarray}
	i\D S_F(x,x')=\frac1{\sqrt{-g}}\delta^{(6)}(x-x')I \ , 
\end{eqnarray}
and can be written in terms of the bispinor $\cal{G}_F$ by
\begin{eqnarray}
\label{S}
	S_F(x,x')=i\D{\cal{G}}_F(x',x) \ , 
\end{eqnarray}
where now ${\cal{G}}_F(x',x)$ obeys the $4\times 4$ differential equation
\begin{eqnarray}
{\bar{\cal{K}}}{\cal {G}}_F(x,x') = -{1\over\sqrt{-g}} \delta^{(6)} (x-x')I\ ,
\label{GF}
\end{eqnarray}
with
\begin{eqnarray}
	{\bar{\cal{K}}}=-\partial_t^2+\alpha^2\left(\partial_r^2+\frac2r\partial_r\right)-\frac{{\vec{L}}^2}{r^2}+\partial_x^2+\partial_y^2-\frac{(1- \alpha)}{r^2}\left(1+{\vec{\Sigma}}\cdot{\vec{L}}\right) \ .
\end{eqnarray}
                     
The vacuum average value for the energy-momentum tensor can be expressed in terms of the Euclidean Green function. It is related with the ordinary Feynman Green function \cite{BD} by the relation  ${\cal G}_E(\tau,\vec{r}; \tau', \vec{r'}) = -i {\cal G}_F(x,x')$, where $t=i\tau$. In the following we shall consider the Euclidean Green function. 

In order to find a solution for the bispinor ${\cal G}_E(x,x')$, we shall obtain the solution for the eigenvalue equation 
\begin{eqnarray}
{\bar{\cal K}}_E \Phi_\lambda (x) = -\lambda^2 \Phi_\lambda
(x)\ ,
\label{Eigenfunction}
\end{eqnarray}
with $\lambda^2 \geq 0$, so we can write 
\begin{eqnarray}
{\cal G}_E(x,x') = \sum_\lambda {\Phi_\lambda (x) \Phi_\lambda^\dagger(x') \over \lambda^2}\ . 
\end{eqnarray}
Due to fact that our operator (\ref{K0}) is self-adjoint, the set of its eigenfunctions constitutes a basis for the Hilbert space associated with four-component spinors. Moreover, because operator ${\bar{\cal K}}_E$ is a parity even operator, its eigenfunctions present a defined parity, so the normalized
eigenfunctions can be written as:
\begin{eqnarray}
\label{Phi}
\Phi_\lambda^{(\sigma)}(x)&=&\frac{e^{-ikx}}{(2\pi)^{3/2}}{\sqrt{\frac{\alpha p}{r}}}
\left(\begin{array}{c}
J_{\nu_{\sigma}}(pr)\varphi^{(\sigma)}_{j,m_j}(\theta,\phi) \\
in_{\sigma }J_{\nu_{\sigma}+n_\sigma}(pr)\hat{r}\cdot\vec{\sigma}\varphi^{(\sigma)}_{j,m_j}(\theta,\phi)
\end{array}\right) \ , \\
\lambda^2&=&k^2+\alpha^2p^2 \ , 
\end{eqnarray}
where $kx={\bar{\eta}}_{ab}k^a x^b=k_0\tau+k_xx+k_yy$, and $J_\nu$ represents the cylindrical Bessel function of order
\begin{eqnarray}
	\nu_\sigma=\frac{j+1/2}\alpha-\frac{n_\sigma}2 \ , \ \mathrm{with} \ n_\sigma=(-1)^\sigma \ , \  \sigma=0, \ 1 \ .
\end{eqnarray}
These functions are specified  by the set of quantum number $(\sigma, \ k^a, \ p, \ j, \ m_j)$, where $k^a\in(-\infty, \ \infty)$, $p\in[0, \ \infty)$, $j=1/2, \ 3/2, \ ...$ denotes the value of the total angular quantum number, $m_j=-j, \ ... ,\ j$  determines its projection. $\sigma$ specifies two types of eigenfunctions with different parities corresponding to $l=j-n_\sigma/2$ being $l$ the orbital quantum number. In (\ref{Phi}), $\varphi^{(\sigma)}_{j,m_j}$ are the spinor spherical harmonics which are eigenfunctions of the operators
$\vec{L}^2$ and $\vec{\sigma}\cdot \vec{L}$ as shown below: 
\begin{eqnarray}
\vec{L}^2\varphi^{(\sigma)}_{j,m_j}&=&l(l+1)\varphi^{(\sigma)}_{j,m_j}\ , \\
\vec{\sigma}\cdot\vec{L} \varphi^{(\sigma)}_{j,m_j}&=&-(1+\kappa^{(\sigma)}) \varphi^{(\sigma)}_{j,m_j}\ , \label{Kappa}
\end{eqnarray} 
with $\kappa^{(0)}=-(l+1)=-(j + 1/2)$ and $\kappa^{(1)}=l=j+1/2$. Explicit form of above standard function are given in Ref. \cite{B-D}, for example. 

Although we have developed this formalism for a six-dimensional spacetime, it can be adapted to a four and five dimensional space. The reason resides in the representations for the Dirac matrices in these dimensions. For four dimensions a irreducible representation for the flat Dirac matrices is the well known $4\times 4$ ones, so we may use $\sigma^M\equiv\gamma^\mu$. As a consequence, in the corresponding analysis we have to discard the derivative with respect to the coordinates $x$ and $y$ in the operator $\cal{K}$. For a five dimensions a possible representation for the flat Dirac matrices is also $4\times 4$ which can be given by $\sigma^M\equiv(\gamma^\mu, \ i\gamma_5)$. In this case the coordinate $y$ should be discarded. So on basis of these arguments it is possible to generalize the eigenfunction of the operator ${\bar{\cal{K}}}_E$ as:
\begin{eqnarray}
\label{Phi-n}
\Phi_\lambda^{(\sigma)}(x)&=&\frac{e^{-ikx}}{(2\pi)^{n/2}}{\sqrt{\frac{\alpha p}{r}}}
\left(\begin{array}{c}
J_{\nu_{\sigma}}(pr)\varphi^{(\sigma)}_{j,m_j}(\theta,\phi) \\
in_{\sigma }J_{\nu_{\sigma}+n_\sigma}(pr)\hat{r}\cdot\vec{\sigma}\varphi^{(\sigma)}_{j,m_j}(\theta,\phi)
\end{array}\right) \ , 
\end{eqnarray}
with $n=1, \ 2, \ 3$. From now on, we shall use the above eigenfunction and make applications for specific values attributed to $n$.

Now we are in position to obtain the bispinor ${\cal{G}}_E$, which is given by
\begin{eqnarray}
{\cal G}_E(x,x') =\int_0^\infty \  ds\int d^nk \int_0^\infty\  dp \sum_{\sigma,j,m_j}\Phi_\lambda^{(\sigma)}(x)\Phi_\lambda^{(\sigma)\dagger}(x') \  e^{-s\lambda^2} \ .
\label{GE}
\end{eqnarray}
Finally substituting (\ref{Phi-n}) into (\ref{GE}) with $\lambda^2=k^2+\alpha^2p^2$, we obtain with the help of \cite{Grad} a closed expression to the Euclidean Green function:
\begin{eqnarray}
\label{GE1}
{\cal{G}}_E(x',x)=\frac{1}{(2\pi rr')^{\frac{n+1}2}}\left(-\frac{\alpha^2}{\sinh u}\right)^{\frac{n-1}2}
\sum_{j,m_j}\left( 
\begin{array}{cc}
F_{j,m_j}(\cosh u, \Omega, \Omega')&0\\
0&F_{j,m_j}(\cosh u, \Omega, \Omega') 
\end{array} \right)  \ , \nonumber\\ 
\end{eqnarray}  
where
\begin{eqnarray}
\label{F}
F_{j,m_j}(\cosh u, \Omega, \Omega')=Q^{\frac{n-1}2}_{\nu_0-1/2}(\cosh u)C_{j,m_j}^{(0)}(\Omega,\Omega')+Q^{\frac{n-1}2}_{\nu_1-1/2} (\cosh u)C_{j,m_j}^{(1)}(\Omega,\Omega')	\ ,
\end{eqnarray}
being 
\begin{eqnarray}
	\cosh u=\frac{r^2+r'^2+\alpha^2(\Delta x)^2}{2rr'}\geq 1 \ , \ (\Delta x)^2={\bar{\eta}}_{ab}(x-x')^a(x-x')^b \ ,
\end{eqnarray}
and $C_{j,m_j}^{(\sigma)}(\Omega,\Omega')=\varphi^{(\sigma)}_{j,m_j}(\theta,\phi)\varphi^{(\sigma)\dagger}_{j,m_j}(\theta',\phi')$ a $2\times 2$ matrix. $Q^\lambda_\nu$ is the associated Legendre function. We can express this function in terms of hypergeometric functions \cite{Grad} by:
\begin{eqnarray}
Q^\lambda_\nu(\cosh u)&=&e^{i\lambda\pi}2^\lambda{\sqrt{\pi}} \ \frac{\Gamma(\nu+\lambda+1)}{\Gamma(\nu+3/2)} \frac{e^{-(\nu+\lambda+1)u}}{(1-e^{-2u})^{\lambda+1/2}}(\sinh u)^\lambda\times \nonumber\\
&&F\left(\lambda+1/2 \ , \ -\lambda+1/2 \ ; \ \nu+3/2 \ ; \ \frac1{1-e^{2u}}\right) \ .
\end{eqnarray}
In this analysis we identify the parameter $\nu$ with $\nu_\sigma-1/2$ and take $\lambda=(n-1)/2$. So the relevant hypergeometric function is
\begin{eqnarray}
	F\left(\frac n2 \ , \ -\frac{n-2}2 \ ; \ \nu_\sigma+1 \ ; \ \frac1{1-e^{2u}}\right) \nonumber \ .
\end{eqnarray}
If $n$ is a even number, this function becomes a polynomial of degree $\frac{n-2}2$; however being $n$ an odd number, this function is an infinite series. 

Taking $\alpha=1$, $\nu_0=j=l+1/2$ and $\nu_1=j+1=l+1/2$; however for the first index the angular quantum number $l$ assumes crescent integer number starting from zero, i.e., $l= \ 0 , \ 1, \ 2, ...$, while for the second index $l=\ 1, \ 2, \ ...$. For $l=0$ the only term that contributes to (\ref{GE1}) in the summation is $Q^{\frac{n-1}2}_{0}(C^{(0)}_{1/2,1/2}+C^{(0)}_{1/2,-1/2})$. For $l\geq 1$ both associated Legendre functions contribute, $\sum_{l\geq 1}Q^{\frac{n-1}2}_{l}\sum_{l,m_l}(C_{l,m_l}^{(0)}+C_{l,m_l}^{(1)})$. Using the explicit expressions for the spinor spherical harmonics, and using the addition theorem for the spherical harmonics, it is possible to express (\ref{GE1}) by:
\begin{eqnarray}
{\cal{G}}_E(x',x)=\frac{1}{(2\pi rr')^{\frac{n+1}2}}\frac1{4\pi}\left(-\frac1{\sinh u}\right)^{\frac{n-1}2}\sum_{l\geq 0}	(2l+1) Q^{\frac{n-1}2}_l(\cosh u)P_l(\cos\gamma)I \ .
\end{eqnarray}
Now taking $n=1, \, 2, \ 3$ and using the sum of Legendre functions and polynomials \cite{Grad}, we obtain standard expressions for the Green function. So:
\begin{itemize}
\item For $n=1$, we have
\begin{eqnarray}
	{\cal{G}}_E(x',x)=\frac1{4\pi^2}\frac1{(x'-x)^2}I \ .
\end{eqnarray}
\item For $n=2$, we have
\begin{eqnarray}
	{\cal{G}}_E(x',x)=\frac1{8\pi^2}\frac1{(x'-x)^3}I \ .
\end{eqnarray}
\item For $n=3$, we have
\begin{eqnarray}
	{\cal{G}}_E(x',x)=\frac1{4\pi^3}\frac1{(x'-x)^4}I \ .
\end{eqnarray}
\end{itemize}
 
After this application of our formalism, let us return to higher dimensional global monopole spacetime. In this space the fermionic Green function, $\cal{S}_F$, can be given by applying the Dirac operator on the bispinor (\ref{GE1}) according to (\ref{S}).

\section{Vacuum Average of the Energy-Momentum Tensor} 
In this section we shall calculate in a explicit way, the renormalized vacuum expectation value (VEV) of the energy-momentum tensor, $\langle T^A_B \rangle_{Ren.}$. Because the metric tensor does not depend of any dimensional parameter, and also because we are working with natural units system, we can infer that the VEV of the energy-momentum tensor depends only on the radial coordinate $r$. Moreover, due to the nonvanishing of Riemann and Ricci tensor, and scalar curvature of the spacetime, this VEV also depends on the arbitrary mass scale, $\mu$, introduced by the renormalization prescription. So, by dimensional analysis it is expected that 
\begin{eqnarray}
\label{T}
	\langle T^A_B \rangle_{Ren.}=\frac1{(\sqrt{4\pi}r)^{n+3}}\left(F^A_B+G^A_B\ln(\mu r/\alpha)\right) \ ,
\end{eqnarray}
where the tensors $F^A_B$ and $G^A_B$ depends only on the parameter $\alpha$. Because the presence of the arbitrary cutoff scale, $\mu$, there is an ambiguity in the definition of (\ref{T}). Finally for a spacetime of odd dimension, i.e., for a even value of $n$, $G^A_B=0$. Obviously the tensors are diagonal and due to the spherical symmetry of the problem we should have $F^\theta_\theta=F^\phi_\phi$ and $G^\theta_\theta=G^\phi_\phi$. 

The renormalized VEV of the energy-momentum tensor must be conserved,
\begin{eqnarray}
\label{CC}
	\nabla_A\langle T^A_B \rangle_{Ren.}=0 \ ,
\end{eqnarray}
and provide the correct trace anomaly for spacetime of even dimension \cite{Chr}:
\begin{eqnarray}
\label{Tr}
	\langle T^A_A \rangle_{Ren.}=\frac1{(4\pi)^{\frac{n+3}2}}Tr\left(a_{(n+3)/2}\right)=\frac{4T}{(\sqrt{4\pi}r)^{n+3}} \ .
\end{eqnarray}

As we shall see, taking into account these informations, it is possible to express all components $F^A_B$ and $G^A_B$ in terms of the zero-zero ones, $F^0_0$ and $G^0_0$, and the trace $T$

Using the point-splitting procedure \cite{BD}, the VEV of the energy-momentum tensor for this four-component spinor Feynman propagator, compatible with the eight-component one, has the following form:
\begin{eqnarray}
\label{EM}
	\langle T_{AB}(x)\rangle=\frac14\lim_{x'\to x}Tr\left[{\tilde\sigma}_A(\nabla_B-\nabla_{B'})+{\tilde\sigma}_B(\nabla_A-\nabla_{A'})\right]S_F(x,x') \ .
\end{eqnarray}
Because the dependence of the fermionic Green function on the time variable, the zero-zero component of the energy-momentum tensor reads:
\begin{eqnarray}
\langle T_{00}(x)\rangle=\lim_{x'\to x}Tr \ \gamma_0\partial_0 S_F(x,x') \ , 
\end{eqnarray}
which can be expressed by 
\begin{eqnarray}
\label{T00}
	\langle T_{00}(x)\rangle=-i\lim_{x'\to x}\partial_t^2 \ Tr{\cal{G}}_F(x',x)=-\lim_{x'\to x}\partial_\tau^2 \ Tr {\cal{G}}_E(x,x') \ .	
\end{eqnarray}
In the obtainment of the above expression we have first taken in the Green function, (\ref{GE1}) and (\ref{F}), the coincidence limit of the angular variable, $\Omega=\Omega'$, and sum over $m_j$. This makes this function proportional to the unit matrix $I$. So, only the term with time derivative in (\ref{DO}) provides a nonvanishing contribution for the zero-zero component of the energy-momentum tensor. 

Now after these brief comments about general properties of the VEV of the energy-momentum tensor, we shall start the explicit calculation of this quantity, specifying the values for $n$. 

\subsection{Case $n=1$}
For $n=1$, the bulk spacetime corresponds to the four-dimensional global monopole spacetime \cite{BV}. For this case the calculations of the vacuum polarization effects associated with massless two-component spinor field on this spacetime has been developed in \cite{Mello5} long time ago. More recently some Casimir densities associated with four-component massive fermionic field obeying the MIT bag boundary condition on, one spherical spherical shell, and two concentric spherical shells, in the global monopole spacetime have been analyzed in \cite{Mello6} and \cite{Mello7}, respectively. 

\subsection{Case $n=2$}
\label{n2}
The case $n=2$ is a new one, it corresponds to a two-dimensional Minkowiski brane with a global monopole on the transverse sub-manifold. For this case the associated Legendre function in (\ref{F}) assumes a very simple expression
\begin{eqnarray}
	Q_{\nu_\sigma-1/2}^{1/2}(\cosh u)=i{\sqrt\frac\pi2}\frac{e^{-\nu_\sigma}}{{\sqrt{\sinh u}}} \ .
\end{eqnarray}
 
Taking $r=r'$ and $\Omega=\Omega'$ into (\ref{GE1}), summing over $m_j$ and using the above expression for the Legendre function, it is possible to develop the sum over $j$, which is a geometric series, we obtain a closed formula for the Euclidean Green function:
\begin{eqnarray}
\label{GE2}
{\cal{G}}_E(x',x)=\frac\alpha{64\pi^2r^3}\frac1{\sinh(u/2)}\frac1{\sinh^2(u/2\alpha)}I \ , 	
\end{eqnarray}
with $u=2{\rm arcsinh}(\alpha\Delta x/2r)$. 

The above Green function is divergent in the coincidence limit, $\Delta x\to 0$. We can verify its singular behavior by expanding (\ref{GE2}) in power series of $\Delta x$:
\begin{eqnarray}
\label{GE-3}
	{\cal{G}}_E(x',x)&=&\left[\frac1{8\pi^2}\frac1{(\Delta x)^3}-\frac{1-\alpha^2}{96\pi^2r^2}\frac1{(\Delta x)}+\frac{1-\alpha^4}{1920\pi^2r^4}(\Delta x)
\right.\nonumber\\
&+&\left.\frac{31\alpha^6-21\alpha^2-10}{483840\pi^2r^6}(\Delta x)^3+O((\Delta x)^5)\right]I \ .
\end{eqnarray}

In order to obtain a finite and well defined result to the VEV of the zero-zero component of the energy-momentum tensor, we must extract all divergent terms in the coincidence limit after applying the second time derivative. In order to do that in a manifest form, we subtract from the Green function the Hadamard one. In \cite{Chr} is given, for any dimensional spacetime, the general formal expression for the Hadamard function. For a five-dimensional spacetime it reads
\begin{eqnarray}
G_H(x',x)=\frac{\Delta^{1/2}(x',x)}{16\pi^2\sqrt{2}}\frac1{\sigma^{3/2}(x',x)}\left[a_0(x,',x)+a_1(x',x)\sigma(x',x)-a_2(x',x)\sigma^2(x',x) \right] \ .
\end{eqnarray}
For this manifold, $\Delta(x',x)$, the Van Vleck-Morette determinant, and the coefficients $a_i(x)$, are given below:
\begin{eqnarray}
\label{Delta}
	\Delta=1 \ , \ a_0=I \ , \ a_1=-\frac{1-\alpha^2}{6r^2}I \ \ {\rm and} \ a_2=-\frac{1-\alpha^4}{60r^4}I \ .
\end{eqnarray}
The one-half of the geodesic distance for $r'=r$ and $\Omega'=\Omega$, is $\sigma(x',x)=\frac{(\Delta x)^2}{2}$. So, we can see that the singular behavior for the Green function, Eq. (\ref{GE-3}), after taking its second time derivative has the same structure as given by the Hadamard function
\begin{eqnarray}
	G_H(x',x)=\left[\frac1{8\pi^2}\frac1{(\Delta x)^3}-\frac{1-\alpha^2}{96\pi^2r^2}\frac1{(\Delta x)}+\frac{1-\alpha^4}{1920\pi^2r^4}(\Delta x) 
\right]I \ .
\end{eqnarray}

So, on basis of this fact we can see that the renormalized VEV of zero-zero component of the energy-momentum vanishes, i.e., 
\begin{eqnarray}
\langle T_0^0(x)\rangle_{Ren.}=\lim_{x'\to x}\partial_\tau^2 \ Tr \left[{\cal{G}}_E(x',x)-G_H(x',x)\right]= 0 \ . 
\end{eqnarray}

For an odd dimensional spacetime there is no trace anomaly, i.e., $\langle T^A_A \rangle_{Ren.}=0$.  Writing $F^A_B=(F^0_0, \ F^r_r, \ F^\theta_\theta , \ F^\phi_\phi, \ F^x_x)$, we have that the sum of all these components vanishes; moreover, the geometric structure of the brane section of this five-dimensional spacetime is Minkowiski-type; consequently the Green function and the Hadamard one depend on the variables on the two dimensional brane by $\Delta x^2=-(\Delta x^0)^2+(\Delta x^4)^2=(\Delta\tau)^2+(\Delta x^4)^2$. By the definition of the VEV of the energy-momentum tensor, Eq. (\ref{EM}), we have
\begin{eqnarray}
	\langle T_4^4(x)\rangle=\lim_{x'\to x}{\rm Tr}\partial_x^2{\cal{G}}_E(x,x') \ .
\end{eqnarray}
So, we can infer that $\langle T_0^0(x)\rangle_{Ren.}=\langle T_4^4(x)\rangle_{Ren.}$. In this way we have $F^0_0=F^x_x$, being both zero. The conservation condition, $\nabla_A\langle T^A_r \rangle_{Ren.}=0$, provides $3F^r_r= -F^\theta_\theta-F^\phi_\phi$, and $\nabla_A\langle T^A_\theta \rangle_{Ren.}=0$,  $F^\theta_\theta=F^\phi_\phi$. So on basis of all these informations we conclude that all components of the tensor $F^A_B$ are zero, consequently,
\begin{eqnarray}
	\langle T^A_A \rangle_{Ren.}=0 \ .
\end{eqnarray}

\subsection{Case $n=3$}
\label{n3}
The case $n=3$ corresponds to a flat three-dimensional brane transverse to a three-dimensional global monopole space. The respective Euclidean Green function is given in terms of an infinite sum of the associated Legendre function $Q_{\nu_\sigma-1/2}^1$ with $C^{(\sigma)}_{j,m_j}$. Taking the angular coincide limit, $\Omega=\Omega'$, the sum $S_0$ below, given in (\ref{GE1}), can be expressed as:
\begin{eqnarray}
\label{S0}
	S_0=\sum_{j,m_j}Q_{\nu_0-1/2}^1(\cosh u)C^{(0)}_{j,m_j}(\Omega,\Omega)=\frac1{4\pi}\sum_{l\geq 0}(l+1)Q_{\frac{l+1}\alpha-1}^1(\cosh u)I_{(2)} 
\end{eqnarray}
and for $S_1$
\begin{eqnarray}
\label{S1}
	S_1=\sum_{j,m_j}Q_{\nu_1-1/2}^1(\cosh u)C^{(1)}_{j,m_j}(\Omega,\Omega)=\frac1{4\pi}\sum_{l\geq 1}lQ_{\frac{l}\alpha}^1(\cosh u)I_{(2)} \ .
\end{eqnarray}
Using $Q_\nu^1(z)=(z^2-1)^{1/2}\frac{dQ_\nu(z)}{dz}$, and the integral representation below for the Legendre function $Q_\nu$
\begin{eqnarray}
\label{Q}
Q_{\nu}(\cosh u)=\frac1{\sqrt{2}}\int_u^\infty \ dt \ \frac{e^{-(\nu+1/2) t}}{\sqrt{\cosh t- \cosh u}} \ ,
\end{eqnarray}
it is possible to develop the sums over $l$ in (\ref{S0}) and (\ref{S1}) getting:
\begin{eqnarray}
	S_0(\cosh u)=\frac{I_{(2)}\sinh u}{16\pi\sqrt{2}}\frac{d}{dz}\left[\int_{\rm arccosh z}^\infty\frac{dt \ e^{t/2}}{\sqrt{\cosh t-z}}\frac1{\sinh^2(t/2\alpha)}\right]{\mid_{z=\cosh u}} 
	\end{eqnarray}
	and
	\begin{eqnarray}
	S_1(\cosh u)=\frac{I_{(2)}\sinh u}{16\pi\sqrt{2}}\frac{d}{dz}\left[\int_{\rm arccosh z}^\infty\frac{dt \ e^{-t/2}}{\sqrt{\cosh t-z}}\frac1{\sinh^2(t/2\alpha)}\right]{\mid_{z=\cosh u}} \ .
\end{eqnarray}

Substituting the sum $S_0+S_1$ into (\ref{GE1}), we can express the Euclidean Green function by
\begin{eqnarray}
\label{GE3}
	{\cal{G}}_E(x',x)=-\frac I{32\pi^3}\frac{\alpha^2}{(r'r)^2}\frac{d}{dz}\left[\int_{\sqrt{z-1}}^\infty\frac{dy}{\sqrt{y^2+1-z}}\frac1{\sinh^2\left(\frac{{\rm arcsinh(y/\sqrt{2})}}\alpha\right)}\right]{\mid_{z=\cosh u}} \ .
\end{eqnarray}
where we have introduced a new variable $t:=2 \ {\rm arcsinh(y/\sqrt{2})}$. 

The VEV of the zero-zero component of the energy-momentum tensor can be formally given by substituting (\ref{GE3}) into (\ref{T00}). Because this procedure provides a divergent result, we must renormalize it by extracting all the divergent terms. In what follows we shall develop a procedure to extract the divercencies in a manifest form. Let us consider the integral inside the bracket of (\ref{GE3}) and write it as:
\begin{eqnarray}
	I_\alpha(z)&=&I_1(z)+I_2(z)=\int_{\sqrt{z-1}}^1\frac{dy}{\sqrt{y^2+1-z}}\frac1{\sinh^2\left(\frac{{\rm arcsinh(y/\sqrt{2})}}\alpha\right)}\nonumber\\
&+& \int_1^\infty\frac{dy}{\sqrt{y^2+1-z}}\frac1{\sinh^2\left(\frac{{\rm arcsinh(y/\sqrt{2})}}\alpha\right)} \ .
\end{eqnarray}
Subtracting and adding into the integrand of $I_1$ the first four terms of power series of the function
\begin{eqnarray}
\label{SH}
	\frac1{\sinh^2\left(\frac{{\rm arcsinh(y/\sqrt{2})}}\alpha\right)}=\frac{2\alpha^2}{y^2}+\frac{(\alpha^2-1)}{3}-\frac{(\alpha^4-1)}{30\alpha^2}y^2 + \frac{(31\alpha^6-21\alpha^2-10)}{3780\alpha^4}y^4 + \ ... 
\end{eqnarray}
we get
\begin{eqnarray}
	I_1(z)&=&I_1^{fin}(z)+I_1^{sing}(z) \ ,
\end{eqnarray}
with
\begin{eqnarray}
	I_1^{fin}(z)&=&\int_{\sqrt{z-1}}^1\frac{dy}{\sqrt{y^2+1-z}}\left[\frac1{\sinh^2\left(\frac{{\rm arcsinh(y/\sqrt{2})}}\alpha\right)}-\frac{2\alpha^2}{y^2}-\frac{(\alpha^2-1)}{3}+\frac{(\alpha^4-1)}{30\alpha^2}y^2\right.\nonumber\\
&-&\left.\frac{(31\alpha^6-21\alpha^2-10)}{3780\alpha^4}y^4\right]
\end{eqnarray}	
and
\begin{eqnarray}
I_1^{sing}(z)=\int_{\sqrt{z-1}}^1\frac{dy}{\sqrt{y^2+1-z}}\left[\frac{2\alpha^2}{y^2}+\frac{(\alpha^2-1)}{3}-\frac{(\alpha^4-1)}{30\alpha^2}y^2 + \frac{(31\alpha^6-21\alpha^2-10)}{3780\alpha^4}y^4\right] \ .
\end{eqnarray}

As we shall see, $I_1^{fin}$ together with $I_2$, provide a finite contribution to the VEV. All the divergences are contained into $I_1^{sing}$.

Taking the derivative of $I_1^{sing}$ with respect to $z$ and substituting $z=\cosh u=1+\frac{\alpha^2\sigma}{r^2}$, with $\sigma=\frac{(\Delta x)^2}2$, we expand the result in power of  which survives after taking the second derivative and the coincidence limit. Doing this procedure we have:
\begin{eqnarray}	
\frac{d_1^{sing}(\sigma)}{dz}&=&-\frac{2r^4}{\alpha^2\sigma^2}-\frac{{\bar{a}}_1r^2}{2\alpha^2\sigma}+ \frac{{\bar{a}}_2}{4}\ln\left(\frac{\sigma\alpha^2} {4r^2}\right)+\nonumber\\
&&\frac{\left(-4\alpha^2-10{\bar{a}}_3+6{\bar{a}}_2-3{\bar{a}}_1-6{\bar{a}}_3\ln\left(\frac{\sigma\alpha^2}{4r^2}\right)\right)\sigma\alpha^2}{16r^2} + \ ... \ \ ,
\end{eqnarray}
with
\begin{eqnarray}
{\bar{a}}_1=\frac{(\alpha^2-1)}{3} \ , \ {\bar{a}}_2=\frac{(\alpha^4-1)}{30\alpha^2} \ {\rm and} \ {\bar{a}}_3= \frac{(31\alpha^6-21\alpha^2-10)}{3780\alpha^4} \ ,
\end{eqnarray}
being last three coefficients of the expansion (\ref{SH}).

We can see that the contribution to the Euclidean Green function given by $I_1^{sing}$ has the same structure as the Hadamard function for a six-dimensional spacetime,
\begin{eqnarray}
\label{GH3}	G_H(x',x)=\frac{\Delta^{1/2}}{16\pi^3}\left[\frac{a_0}{\sigma^2}+\frac{a_1}{2\sigma}-\frac14\left(a_2-\frac{a_3}2\sigma)\right)\ln\left(\frac{\mu^2\sigma}{4}\right)\right] \ ,
\end{eqnarray}
where $\mu$ is an arbitrary energy scale \cite{Chr}. $\Delta$, $a_0$, $a_1$ and $a_2$ are same as given in (\ref{Delta}) for this six-dimensional global monopole spacetime.

In \cite{Gilkey,JP}, an explicit expression for the coefficient $a_3$ is provided for a general second order differential operator $D^2+X$, $D_M$ being the covariant derivative including gauge field and $X$ an arbitrary scalar function. For this six-dimensional global monopole spacetime, by using the computer program {\it GRTensorII}, we found, considering $X=-\frac14RI$,  
\begin{eqnarray}
	a_3=\frac{31\alpha^6-21\alpha^2-10}{2520r^6}I \ .
\end{eqnarray}

The renormalized bispinor given by
\begin{eqnarray}
{\cal{G}}_{Ren.}(x',x)={\cal{G}}_E(x',x)-G_H(x',x) \ , 
\end{eqnarray}
with
\begin{eqnarray}
	{\cal{G}}_E(x',x)=-I\frac{\alpha^2}{32\pi^3r^4}\left[\frac{d I_1^{sing}(\sigma)}{dz}+\frac{d I_1^{fin}(\sigma)}{dz} + \frac{d I_2(\sigma)}{dz}\right]
\end{eqnarray}
and $G_H(x',x)$ given by (\ref{GH3}), provides
\begin{eqnarray}
\label{GR}
{\cal{G}}_{Ren.}(x',x)&=&I\frac{\alpha^2}{64\pi^3r^4}\left[{\bar{a}}_2-\frac{3{\bar{a}}_3\alpha^2\sigma}{2r^2}\right]\ln\left(\frac{\mu r}{\alpha}\right)- I\frac{[-4\alpha^2-10{\bar{a}}_3+6{\bar{a}}_2-3{\bar{a}}_1]\alpha^4\sigma}{512\pi^3r^6}\nonumber\\
&-&I\frac{\alpha^2}{32\pi^3r^4}\left[\frac{dI_1^{fin}(\sigma)}{dz} + \frac{d I_2(\sigma)}{dz}\right] \ .
\end{eqnarray}

Now we are in position to obtain the renormalized VEV of the zero-zero component of the energy momentum tensor. It is given by:
\begin{eqnarray}
\label{TR}
	\langle T_0^0(x)\rangle_{Ren.}=\lim_{x'\to x}\partial_\tau^2 \ Tr  \ {\cal{G}}_{Ren.}(x',x) \ . 
\end{eqnarray}

Wald in \cite{Wald}, has proved that in order to obtain a energy-momentum tensor which obeys the conservation condition law (\ref{CC}) and correct trace anomaly (\ref{Tr}), an additional contribution must be considered. In a six-dimensional spacetime, this terms reads:
\begin{eqnarray}
\label{AA}
	\frac1{384\pi^3}\delta_A^B{\rm Tr}a_3 \ .
\end{eqnarray}

So on basis on this fact we explicitly present in Appendix \ref{A} the steps needed to obtain $F_0^0$ and $G_0^0$. Particularment with respect to $F_0^0$ we provide an integral expression for it, and for $G^0_0$ a closed expression. Briefly speaking, they are given by substituting (\ref{GR}) into (\ref{TR}) and taking into account (\ref{AA}). These components are:
\begin{eqnarray}
\label{F00}
	F_0^0=-6\alpha^4\int_0^1 \ \frac{dx}{x^5}f_1(x)-6\alpha^4\int_1^\infty \ \frac{dx}{x^5}f_2(x)+6\alpha^4{\bar{a}}_3  
\end{eqnarray}
with
\begin{eqnarray}
f_1(x)&=&\frac1{\sinh^2\left(\frac{{\rm arcsinh(x/\sqrt{2})}}\alpha\right)} -\frac{2\alpha^2}{x^2}-\frac{(\alpha^2-1)}{3}+\frac{(\alpha^4-1)}{30\alpha^2}x^2 \nonumber\\
&-&\frac{(31\alpha^6-21\alpha^2-10)}{3780\alpha^4}x^4 \ ,
\end{eqnarray}
\begin{eqnarray}
f_2(x)=\frac1{\sinh^2\left(\frac{{\rm arcsinh(x/\sqrt{2})}}\alpha\right)} -\frac{2\alpha^2}{x^2}-\frac{(\alpha^2-1)}{3}+\frac{(\alpha^4-1)}{30\alpha^2}x^2 
\end{eqnarray}
and
\begin{eqnarray}
\label{G00}
	G_0^0=-r^6 {\rm Tr}\ {a}_3 =-\frac{31\alpha^6-21\alpha^2-10}{630}\ .
\end{eqnarray}
In (\ref{F00}) we have used the fact that $(r^6/6) {\rm Tr}{a}_3=\alpha^4{\bar{a}}_3$.

\begin{figure}[tbph]
\begin{center}
\begin{tabular}{cc}
\epsfig{figure=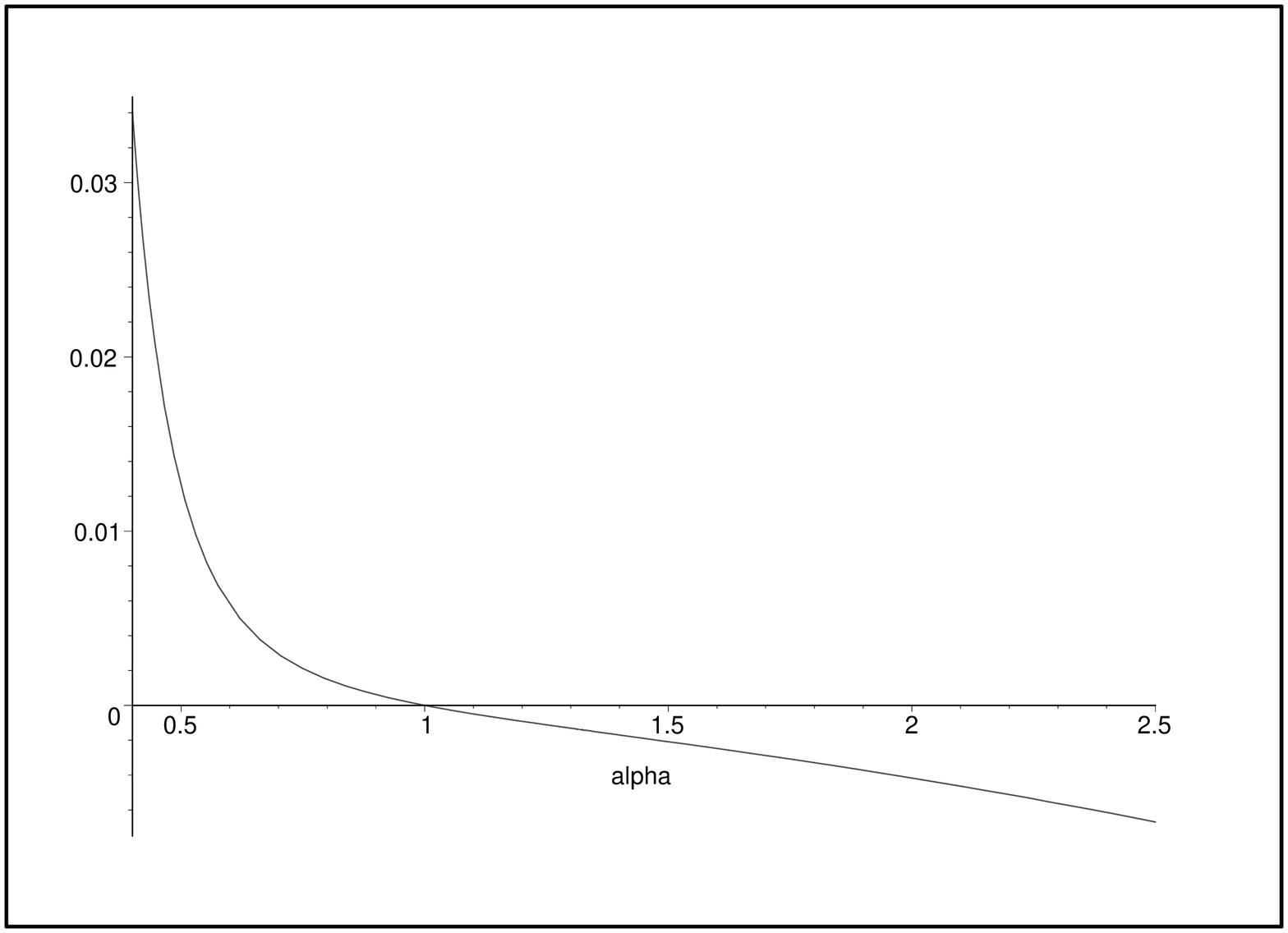, width=7cm, height=6cm,angle=-90} & \quad 
\epsfig{figure=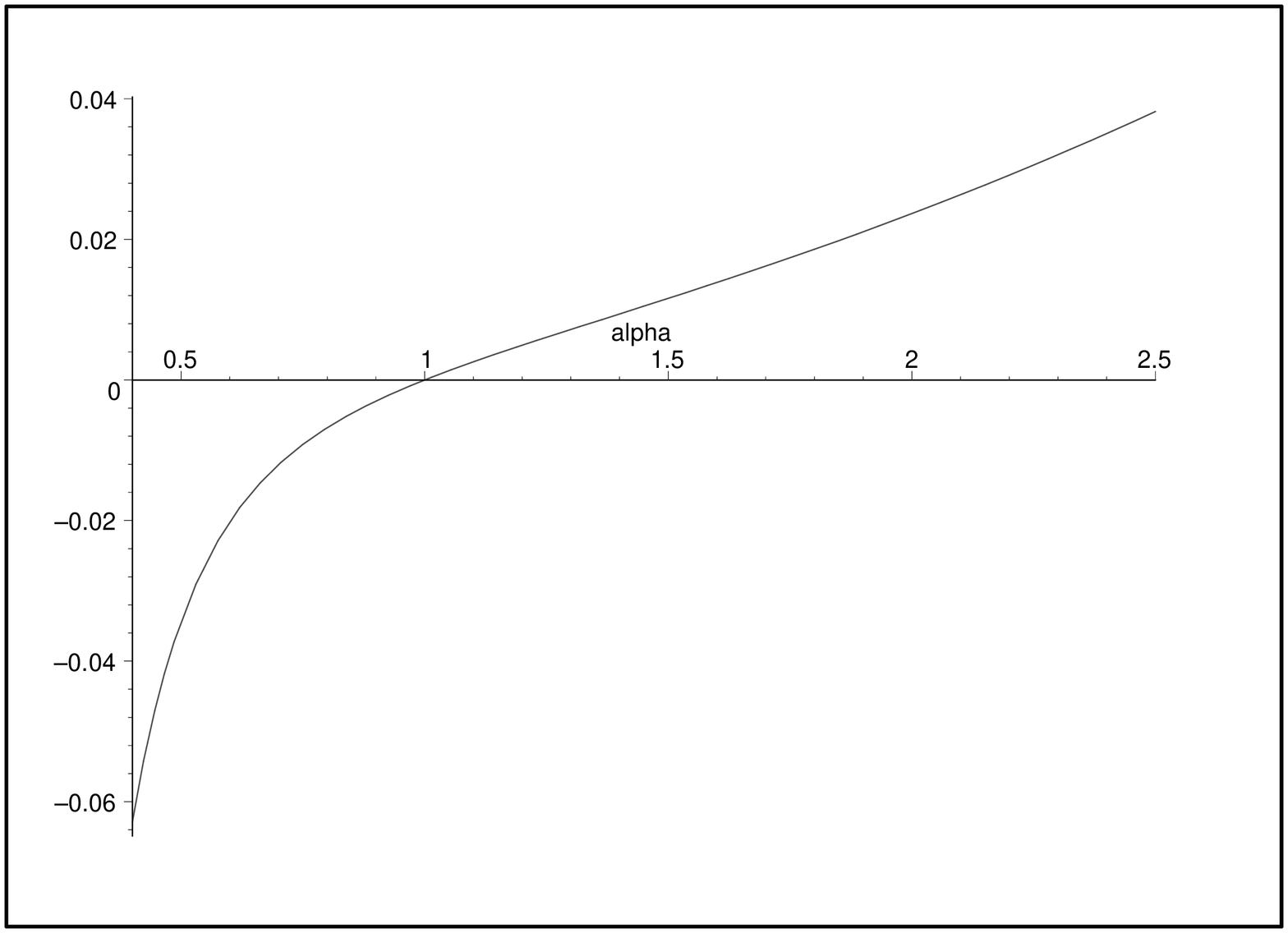, width=7cm, height=6cm,angle=-90}
\end{tabular}
\end{center}
\caption{These graphs represent the dependence of the two integrals in (\ref{F00}) as a function of the parameter $\alpha$ in a specific interval. The left panel corresponds to integral associated with the function $f_1(x)/x^5$, and the right panel corresponds to integral associated with the function $f_2(x)/x^5$.}
\label{fig1}
\end{figure}

Unfortunately it is not possible to provide analitycal results to the integrals above for a general value of $\alpha$. Their dependence on the parameter $\alpha$ can only be provided numerically. Our numerical results to these integrals are exhibited in figure \ref{fig1}. Also it is possible to provide the approximate behavior for the integrals, $I_1=\int_0^1 \frac{dx}{x^5}f_1$ and $I_2=\int_1^\infty\frac{dx}{x^5}f_2$, for specific limit of this parameter:
\begin{itemize}
\item For large solid angle deficit ($\alpha<<1$),
\begin{eqnarray}
	I_1(\alpha)&\approx& -\frac1{378}\frac{\ln(\alpha)}{\alpha^4} \\
	I_2(\alpha)&\approx&-\frac1{60\alpha^2} \ .
\end{eqnarray}
\item For small solid angle deficit ($|\alpha-1|<<1)$,
\begin{eqnarray}
	I_1(\alpha)&\approx&-0.0052(\alpha-1) \\
	I_2(\alpha)&\approx&0.0275(\alpha-1) \ .
\end{eqnarray}
\item For large angle solid excess ($\alpha>>1$),
\begin{eqnarray}
	I_1(\alpha)&\approx&-0.0025\alpha^2 \\
	I_2(\alpha)&\approx&\frac{\alpha^2}{60} \ .
\end{eqnarray}
\end{itemize}

Having found $F^0_0$ and $G_0^0$ the others components of the renormalized VEV of the energy-momentum tensor can be expressed in terms of them. Due to the Minkowiski structure of the brane section of this spacetime, the Green function and also the Hadamard one depend on the variables on the brane by $(\Delta x)^2={\bar{\eta}}_{ab}(x-x')^a(x-x')^b$; on the other hand, by using (\ref{EM}) we can verify that
\begin{eqnarray}
	\langle T_a^a(x)\rangle=\lim_{x'\to x}{\rm Tr} \ \partial_a^2 {\cal{G}}_E(x',x) \ , \  {\rm for} \ a=4, \ 5 \ .
\end{eqnarray}
So we may conclude that the renormalized VEV of these components are equal to zero-zero one. Consequently $F_0^0=F_4^4=F_5^5$ and $G_0^0= G_4^4=G_5^5$. Besides the trace anomaly
\begin{eqnarray}
	\langle T^A_A(x)\rangle_{Ren.}=\frac1{64\pi^3}{\rm Tr}a_3=\frac T{16\pi^3r^6} \ ,
\end{eqnarray}
with
\begin{eqnarray}
	T=\frac{31\alpha^6-21\alpha^2-10}{2520} \ 
\end{eqnarray}
give us $F^A_A=4T$ and $G_A^A=0$. The conservation conditions, $\nabla_A\langle T^A_\theta\rangle_{Ren.}=0$, and $\nabla_A\langle T^A_r\rangle_{Ren.}=0$, provide, respectively,  $T^\theta_\theta=T^\phi_\phi$ and $4F^r_r-G_r^r+2F^\theta_\theta=0$. So after a simple calculation we can express:
\begin{eqnarray}
	F^A_B&=&{\rm diag}\left(F^0_0, \ F_0^0+G_0^0/3-4T/3, \ 8T/3-2F^0_0-G_0^0/6, \right.\nonumber\\
	&&\left. \ 8T/3-2F^0_0-G_0^0/6, \ F^0_0, \ F^0_0 \right) 
\end{eqnarray}
and
\begin{eqnarray}
	G^A_B=G_0^0{\rm diag}\left(1, \  1, \ -2, \ -2, \ 1, \ 1\right) \ .
\end{eqnarray}

\section{Concluding Remarks}
In this paper we have considered the vacuum polarization effect associated with fermionic field induced by a global monopole in the braneworld context. By this scenario our Universe is described by a $(n-1)-$flat brane transverse to a point-like three-dimensional global monopole manifold. Two specific spacetime have been explicitly considered, the cases with $n=2$ and $3$. In fact our first motivation was to consider a six-dimensional bulk. However, because we have considered a defined positive chiral field, the fermionic field can be described by a four-component representation; so, in this way, we could extend the formalism to include a five-dimensional bulk. Due to the fact that the global monopole lives in a three-dimensional manifold, it was possible to express the fermionic eigenfunctions in terms of spinor spherical harmonics, $\varphi^{(\sigma)}_{j,m_j}$. Differently from the scalar case analyzed in \cite{Mello4}, the effective total angular momentum have simple expressions: $\nu_0=(l+1)/\alpha-1/2$ and $\nu_1=l/\alpha+1/2. $\footnote{For the scalar case the effective orbital angular quantum number assumes a simple form only for a curvature coupling parameter $\xi=1/8$.} 

\begin{figure}[tbph]
\begin{center}
\begin{tabular}{cc}
\epsfig{figure=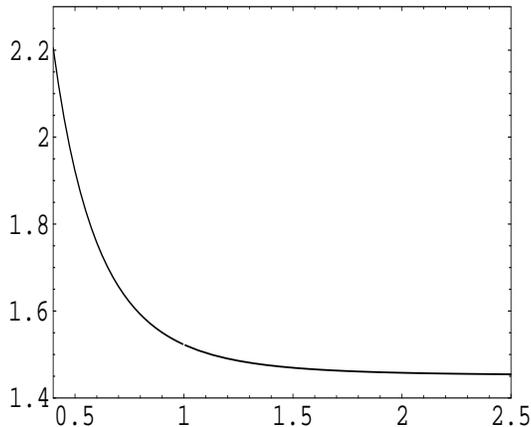, width=7cm, height=6cm} 
\end{tabular}
\end{center}
\caption{This graph represents the dependence of the physical distance $\mu\bar{r}/\alpha$, that $\langle T_0^0(x)\rangle_{Ren.}$ vanishes, with $\alpha$.}
\label{fig2}
\end{figure}

The main objective of this paper was to obtain the renormalized vacuum expectation value of the energy-momentum tensor, $\langle T_A^B\rangle$. In order to do that we have explicitly constructed the fermionic Green function, which was expressed in terms of a linear differential operator acting on a bispinor. Because all components of this tensor can be related by the conservation condition and the correct trace anomaly, we needed only to calculate the zero-zero component of this tensor. By our results we could verify that for a five-dimensional bulk $\langle T_A^B\rangle_{Ren.}=0$. \footnote{A vanishing result for the renormalized vacuum expectation value of the square of the field, $\langle\Phi^2\rangle$, have also been obtained for a scalar field in this five-dimensional bulk in a specific approximated result \cite{Mello4}.} However, for a six-dimensional bulk, a non-vanishing result have been obtained. This result depends on the inverse of the sixth order power of the distance from the point to the monopole's core considered on the brane, and also on the arbitrary energy scale $\mu$:
\begin{eqnarray}
\label{Too}
	\langle T_0^0(x)\rangle_{Ren.}=\frac1{64\pi^3r^6}\left(F^0_0(\alpha)+G_0^0(\alpha)\ln(\mu r/\alpha)\right) \ .
\end{eqnarray}
The expressions found for $F_0^0$ involve a long calculation and some details related with this calculation is presented in Appendix \ref{A}. By our result this component is given in terms of a closed term and two more integral expressions. For these integrals we provided numerically, in Fig. \ref{fig1}, their behavior as function of $\alpha$. Fortunately we were able to provide a closed expression for $G_0^0$, which by its turn depends on the coefficient $a_3$ of the Hadamard function. The general expression for this coefficient, for the fermionic case, presents $43$ terms, so its final result requires a long and careful calculation, even using the specific computer program GRTensorII. Because (\ref{Too}) is given in terms of two distinct contributions, there exist a radius $\bar{r}$ which $\langle T_0^0(x)\rangle_{Ren.}$ vanishes. In figure \ref{fig2} we present the dependence of the dimensionless distance $\mu\bar{r}/\alpha$ as a function of $\alpha$.

Finally we would like to mention that the model analysed here presents monopole as a point-like object having its core on the flat brane; so the influence due to the fermionic quantum field on the brane can be evaluated in the region near the monopole's core. However, by this model the renormalized VEV of the energy-momentum tensor is divergent at the monopole's core. The problem of the singularity of vacuum polarization effects involving a global monopole, and topological deffects in general, can be avoided by considering a more realistic model to the monopole, i.e., considering a inner structure to its core. A simplified model for the monopole core has been first presented in \cite{Lousto}. The vacuum polarization effects due to a massless scalar field in the region outside this model have been investigated in \cite{Jean}. More recently the analysis of vacuum polarization effect associated with quantum bosonic and fermionic fields, in the global monopole spacetime with a general spherically symmetric inner structure, have been developed in \cite{Saha1} and \cite{Saha2}, respectively. In these analysis the asymptotic behavior of the core-induced vacuum densities are investigated at large and small distance from the core, and for small a large angle solid deficit.  

\section*{Acknowledgment} 
The author would like to thank J. Batista Fonseca, for his help with the computer program GRTensorII and to Aram A. Saharian, also to Conselho Nacional de Desenvolvimento Cient\'\i fico e Tecnol\'ogico (CNPq.) for partial financial support, FAPESQ-PB/CNPq. (PRONEX) and FAPES-ES/CNPq. (PRONEX) .

\appendix
\section{Explicit Calculation of $F^0_0$ and $G^0_0$}\label{A}
In order to find the complete expression for (\ref{GR}), it is necessary to obtain the derivative of $I_1^{fin}$ and $I_2$ with respect to $z$. Our first steps in this direction is to introduced a new variable $b=\sqrt{z-1}$. So we may write:
\begin{eqnarray}
	I_1^{fin}(z)=\int_b^1\frac{dx}{\sqrt{x^2-b^2}}f_1(x) \ ,
\end{eqnarray}
with 
\begin{eqnarray}
	f_1(x)&=&\frac1{\sinh^2\left(\frac{{\rm arcsinh(x/\sqrt{2})}}\alpha\right)} -\frac{2\alpha^2}{x^2}-\frac{(\alpha^2-1)}{3}+\frac{(\alpha^4-1)}{30\alpha^2}x^2 \nonumber\\
&-&\frac{(31\alpha^6-21\alpha^2-10)}{3780\alpha^4}x^4  \ .
\end{eqnarray}
Consequently
\begin{eqnarray}
	\frac{I_1^{fin}(\sigma)}{dz}=\frac1{2b}\frac d{db}\int_b^1\frac{dx}{\sqrt{x^2-b^2}}f_1(x) \ .
\end{eqnarray}
Because the integrand of $I_1^{fin}$ is divergent at the point $x=b$, we have to change the variable $x\to bx$  before applying the Leibnitz formula for derivative of an integral. So, this derivative can be written as
\begin{eqnarray}
	\frac{I_1^{fin}(\sigma)}{dz}=\frac12\int_b^1\frac{dx}{\sqrt{x^2-b^2}}g(x)-\frac{f_1(1)}{2\sqrt{1-b^2}} \ ,
\end{eqnarray}
where $g(x)=\frac d{dx}(\frac{f_1(x)}{x})$ and $b=\frac{\alpha\sqrt{\sigma}}r$. The derivative of $I_2$ is simple and reads
\begin{eqnarray}
		\frac{I_2(\sigma)}{dz}=\frac12\int_1^\infty\frac{dx}{(x^2-b^2)^{3/2}}\frac1{\sinh^2\left(\frac{{\rm arcsinh(x/\sqrt{2})}}\alpha\right)} \ .
\end{eqnarray}

Using these results we can express the renormalized Green function, Eq. (\ref{GR}), as:
\begin{eqnarray}
\label{GR1}
{\cal{G}}_{Ren.}(x',x)&=&I\frac{\alpha^2}{64\pi^3r^4}\left[{\bar{a}}_2-\frac{3{\bar{a}}_3\alpha^2\sigma}{2r^2}\right]\ln\left(\frac{\mu r}{\alpha}\right)- I\frac{[-4\alpha^2-10{\bar{a}}_3+6{\bar{a}}_2-3{\bar{a}}_1]\alpha^4\sigma}{512\pi^3r^6}\nonumber\\
&-&I\frac{\alpha^2}{64\pi^3r^4}\left[\int_b^1\frac{dx}{\sqrt{x^2-b^2}}g(x)-\frac{f_1(1)}{\sqrt{1-b^2}}\right] \nonumber\\
&-&I\frac{\alpha^2}{64\pi^3r^4}\int_1^\infty\frac{dx}{(x^2-b^2)^{3/2}}\frac1{\sinh^2\left(\frac{{\rm arcsinh(x/\sqrt{2})}}\alpha\right)} \ .
\end{eqnarray}

The next step, according to (\ref{TR}), is to obtain the second derivative with respect to the Euclidean time. Taking the coincidence limits $x'=x$ and $y'=y$, using the fact that $\partial_\tau^2=(\alpha^2/2r^2)\partial_b^2$, and adopting the same procedure as explained above we can write
\begin{eqnarray}
	\frac{d^2}{db^2}\int_b^1\frac{dx}{\sqrt{x^2-b^2}}g(x)&=&\int_b^1\frac{dx}{\sqrt{x^2-b^2}}\left(g''(x)-\frac{g'(x)}x+\frac{g(x)}{x^2}\right)- \frac{g'(1)}{\sqrt{1-b^2}}\nonumber\\
&-&\frac{g(1)b^2}{(1-b^2)^{3/2}} \ .
\end{eqnarray}
Taking the limit $b\to 0$ we obtain
\begin{eqnarray}
\frac{d^2}{db^2}\int_b^1\frac{dx}{\sqrt{x^2-b^2}}g(x)\to \int_0^1 \ dx \ \frac{g(x)}{x^3} \ .
\end{eqnarray}
We can see that the integral above is finite because for small values of $x$, $g(x)\to -\frac{(289\alpha^8-168\alpha^4-100\alpha^2-21)} {22.680\alpha^6}x^4+O(x^6)$. In the limit $b\to 0$, the other relevant terms in the renormalized Green function above provide
\begin{eqnarray}
	\frac{d^2}{db^2}\frac1{(x-b^2)^{3/2}}&\to&\frac3{x^5} \ , \\
	\frac{d^2}{db^2}\frac1{\sqrt{1-b^2}}&\to&1 \ .
\end{eqnarray}

So taking into account all these results and including the additional term (\ref{AA}) we have:
\begin{eqnarray}
	\langle T_{00}(x)\rangle_{Ren.}&=&-\frac{\alpha^6}{32\pi^3r^6}+\frac{[6{\bar{a}}_2-3{\bar{a}}_1-10{\bar{a}}_3]\alpha^4}{128\pi^3r^6}+\frac{3\alpha^4} {32\pi^3r^6}\int_1^\infty\frac{dx}{x^5}\frac1{\sinh^2\left(\frac{{\rm arcsinh(x/\sqrt{2})}}\alpha\right)}\nonumber\\
	&+&\frac{3\alpha^4} {32\pi^3r^6}\int_0^1\frac{dx}{x^5}f_1(x)+\frac{3{\bar{a}}_3\alpha^4}{32\pi^6r^6}\ln\left(\frac{\mu r}\alpha\right) -\frac1{64\pi^3r^6}\alpha^4{\bar{a}}_3 \ ,
\end{eqnarray}
which can be written as shown below:
\begin{eqnarray}
	\langle T_{00}(x)\rangle_{Ren.}&=&\frac{3\alpha^4} {32\pi^3r^6}\int_0^1\frac{dx}{x^5}f_1(x)+\frac{3\alpha^4} {32\pi^3r^6}\int_1^\infty\frac{dx}{x^5}f_2(x)\nonumber\\
	&-&\frac{3\alpha^4}{32\pi^3r^6}{\bar{a}}_3+\frac{3{\bar{a}}_3\alpha^4}{32\pi^3r^6}\ln\left(\frac{\mu r}\alpha\right) \ .
\end{eqnarray}

\end{document}